\documentclass[twocolumn,floats,floatfix,superscriptaddress,noeprint,aps,pra]{revtex4-1}

\newcommand{\pryso}{$\text{Pr}^{3+}\text{:}\text{Y}_2\text{SiO}_5\:$}
\newcommand{\prysoshort}{Pr:YSO}

\newcommand{\mi}{\mathrm{i}}
\newcommand{\rabi}[1]{\Omega_\text{#1}}
\newcommand{\detuning}[1]{\Delta_\text{#1}}

\def\i{\,\text{i}}
\def\e{\,\text{e}}

\usepackage{amsfonts,amssymb,amsmath}
\usepackage{calc,xcolor-patch}
\usepackage{graphicx}
\usepackage{braket}
\usepackage{siunitx}
\usepackage{bm}
\usepackage[normalem]{ulem}
\usepackage{todonotes}

\DeclareSIUnit{\atpercent}{at.~\percent}

\begin{document}


\title{Experimental demonstration of composite stimulated Raman adiabatic passage}

\author{Alexander Bruns}
\email{alexander.bruns@physik.tu-darmstadt.de}
\affiliation{Institut f{\"u}r Angewandte Physik, Technische Universit{\"a}t Darmstadt, Hochschulstra{\ss}e 6, 64289 Darmstadt, Germany}
\author{Genko T. Genov}
\affiliation{Institut f{\"u}r Angewandte Physik, Technische Universit{\"a}t Darmstadt, Hochschulstra{\ss}e 6, 64289 Darmstadt, Germany}
\author{Marcel Hain}
\affiliation{Institut f{\"u}r Angewandte Physik, Technische Universit{\"a}t Darmstadt, Hochschulstra{\ss}e 6, 64289 Darmstadt, Germany}
\author{Nikolay V. Vitanov}
\affiliation{Department of Physics, St. Kliment Ohridski University of Sofia, 5 James Bourchier Boulevard, 1164 Sofia, Bulgaria}
\author{Thomas Halfmann}
\homepage{http://www.iap.tu-darmstadt.de/nlq}
\affiliation{Institut f{\"u}r Angewandte Physik, Technische Universit{\"a}t Darmstadt, Hochschulstra{\ss}e 6, 64289 Darmstadt, Germany}

\date{\today}

\begin{abstract}	
	We experimentally demonstrate composite stimulated Raman adiabatic passage (CSTIRAP), which combines the concepts of composite pulse sequences and adiabatic passage. The technique is applied for population transfer in a rare-earth doped solid. We compare the performance of CSTIRAP with conventional single and repeated STIRAP, either in the resonant or the highly detuned regime. In the latter case, CSTIRAP improves the peak transfer efficiency and robustness, boosting the transfer efficiency substantially compared to repeated STIRAP. We also propose and demonstrate a universal version of CSTIRAP, which shows improved performance compared to the originally proposed composite version. Our findings pave the way towards new STIRAP applications, which require repeated excitation cycles, e.g., for momentum transfer in atom optics, or dynamical decoupling to invert arbitrary superposition states in quantum memories.
\end{abstract}

\maketitle


\section{Introduction\label{secIntro}}
Efficient techniques to coherently control quantum systems are essential for evolving quantum technologies. A large variety of control techniques aim at efficient quantum state inversion, e.g., for applications in physical chemistry \cite{Brif2010,*Rice2000,*Shapiro2011,*Tannor2007}, nuclear magnetic resonance \cite{Levitt1986}, or quantum information processing \cite{Schmidt-Kaler2003,*Riebe2008,*Monz2009,Pons2007,*Timoney2008,*Piltz2013}. The main requirements for efficient population transfer are high fidelity and robustness against fluctuations in experimental parameters, while maintaining short excitation times.

Resonant two-level techniques, such as resonant $\pi$-pulses, in principle permit high fidelity population inversion at short interaction times, but they usually suffer substantially from inevitable variations in the experimental parameters.
Adiabatic processes, e.g., rapid adiabatic passage (RAP) \cite{Vitanov2001} or stimulated Raman adiabatic passage (STIRAP) \cite{Vitanov2017}, are robust alternatives. STIRAP is among the most established adiabatic control tools for coherent population transfer between quantum states. STIRAP found a multitude of applications in atomic physics, molecular physics, solid-state physics, nonlinear optics, quantum information technology, and many others \cite{Vitanov2017}. Despite its robustness and high efficiency, STIRAP usually requires rather long interaction times and/or high intensities to reach the required adiabaticity for high fidelity population transfer. Moreover, STIRAP is very sensitive with regard to the proper preparation of a pure initial state. This is a severe obstacle, if repeated (cyclic) application of STIRAP is required.
The latter is essential, e.g., in atom optics, when repeated STIRAP is applied to obtain large momentum transfer and beam deflection in real space. The fast drop in efficiency limits the number of possible repetitions and the maximal deflection angle \cite{Vitanov2017}. Moreover, cyclic STIRAP processes also offer potential to adiabatically drive logic operations \cite{Beil2011} or to invert (or dynamically decouple) coherent superposition states in quantum memories.

Various approaches can be employed to speed up adiabatic techniques, e.g., optimal control \cite{Glaser2015} or single-shot shaped pulses \cite{Daems2013,*Van-Damme2017}. These usually rely on pulse shaping in the frequency or time domain, or additional fields to compensate for unwanted diabatic transitions \cite{Vitanov2017}.
Composite pulses are another alternative to improve the fidelity and robustness of coherent excitation processes with the advantage that they do not require compensation of single pulse diabatic losses. They were initially developed and are well established in nuclear magnetic resonance \cite{Levitt1986}. In recent years, they also found their way into quantum optics and quantum information processing \cite{Wesenberg2003,Haffner2008,Torosov2011,*Ivanov2011,*Genov2011,*Genov2013}. Composite pulses drive robust excitation pathways in Hilbert space between an initial and a desired final state. The relative phases of the pulses in a composite sequence serve as control parameters, allowing for compensation against certain experimental imperfections. It is also possible to design universal composite pulses, which compensate against any arbitrary variation of experimental parameters in the excitation process. We theoretically proposed and experimentally demonstrated such universal pulse sequences in previous work, which aimed in particular at dynamical decoupling \cite{Genov2014,Genov2017a}.
It is a promising idea to combine the concepts of composite pulse sequences with adiabatic passage, in order to improve arbitrary properties of the adiabatic excitation processes, e.g., efficiency, bandwidth, or robustness. We already implemented a composite version of RAP (termed composite adiabatic passage), which permits efficient excitation in a two-level system \cite{Torosov2011a,*Schraft2013}. Recently, a combination of composite pulse sequences with STIRAP (which we will term now CSTIRAP), was theoretically proposed for population transfer in a three-state system \cite{Torosov2013}. However, CSTIRAP has not yet been implemented experimentally.

In the following, we report on the proof-of-principle experimental demonstration and thorough systematic investigation of CSTIRAP. Specifically, we apply it for population transfer in a rare-earth ion-doped crystal. In the latter medium, we already implemented conventional STIRAP \cite{Klein2007} and applied cyclic STIRAP sequences for classical information processing \cite{Beil2011}.
We measure the population transfer efficiency and compare the performance of CSTIRAP with conventional and repeated STIRAP in terms of fidelity and robustness. We investigate two versions of CSTIRAP for resonant and detuned excitations, which were already theoretically proposed \cite{Torosov2013}. We also develop and experimentally demonstrate a novel, universal variant of detuned CSTIRAP, with improved performance compared to the originally proposed composite version.

Our work paves the way for applications of CSTIRAP in all fields
where the widely used STIRAP is applicable, i.e., well beyond the
specific experimental implementation presented below.
CSTIRAP offers particular advantages when repeated (cyclic) STIRAP is required,
e.g., in atom optics to obtain large momentum transfer and beam deflection in real space,
or for all-optical spin rephasing and dynamical decoupling
\cite{Rui2015,Serrano2018}. As CSTIRAP does not rely on pulse
shaping, it can also be combined with optimal control or
shortcuts-to-adiabaticity improved versions of STIRAP \cite{Du2016}
and improve their performance even further.

\section{Theoretical background\label{secTheory}}

\subsection{STIRAP\label{subsecSTIRAP}}
\begin{figure}
	\includegraphics[width=\columnwidth]{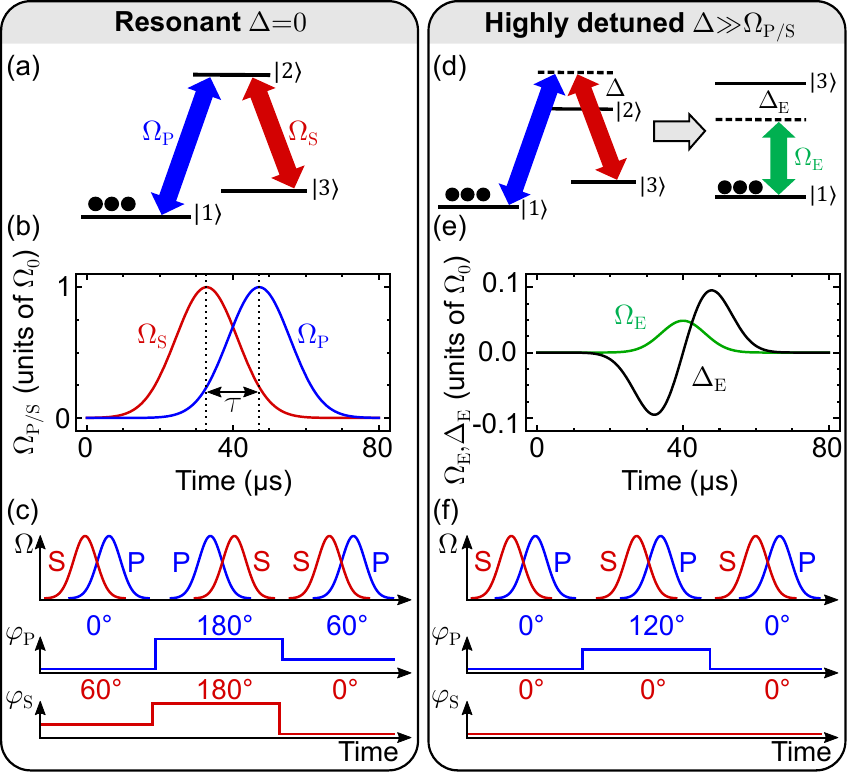}
	\caption{Coupling schemes and (C)STIRAP pulse sequences for resonant (a-c) and highly detuned (d-f) excitation.  Upper row: Basic coupling schemes. Middle row: (b) STIRAP pulse sequence for both excitation regimes. (e) Effective two-photon Rabi frequency  and effective detuning in the Raman-type effective two-state system after adiabatic elimination of the intermediate state. Lower row: Temporal evolution and phases of CSTIRAP sequences (with composite sequences of three pulses as an example). $\varphi_\text{P}$ and $\varphi_\text{S}$ are the phases of the single pump and Stokes pulses.
	}
	\label{figTheoryResVsDetRegime}
\end{figure}
We consider a three-state $\Lambda$-system, e.g., as shown in Fig. \ref{figTheoryResVsDetRegime}. We aim for coherent population transfer from the initial state $\ket{1}$ to the final state $\ket{3}$, mediated via couplings to an intermediate state $\ket{2}$ by two laser fields (pump and Stokes). Initially all population of the system is assumed to be in state $\ket{1}$. The single-photon detunings of the driving laser fields from the corresponding resonances are defined as $\Delta_\text{P}=\omega_\text{P}-\omega_\text{12}$ and $\Delta_\text{S}=\omega_\text{S}-\omega_\text{32}$, which also determine the two-photon detuning $\delta=\Delta_\text{P}-\Delta_\text{S}$. The coupling strengths are given by the Rabi frequencies $\rabi{P}(t)=-\mu_{12}\mathcal{E}_\text{P}(t)/\hbar$ and $\rabi{S}(t)=-\mu_{23}\mathcal{E}_\text{S}(t)/\hbar$ \cite{Shore2011}. Here, $\mu_{ij}$ are the transition dipole moments and $\mathcal{E}_\text{P/S}(t)$ are the time-varying envelopes of the electric fields. On two-photon resonance we have $\delta=0$, hence $\detuning{P}=\detuning{S}\equiv \Delta$, and the system dynamics are described by the Hamiltonian in the rotating wave approximation
\begin{align}
\label{eqHamiltonianRWA}
\hat{H}_\text{RWA}(t)=
\frac{\hbar}{2}
\begin{pmatrix}
0 & \rabi{P}(t) & 0 \\
\rabi{P}(t) & 2\Delta & \rabi{S}(t) \\
0 & \rabi{S}(t) & 0
\end{pmatrix}.
\end{align}
The dynamics of STIRAP are best understood in the adiabatic basis, defined by the instantaneous eigenstates \cite{Vitanov2017}
\begin{subequations}
	\label{eqAdiabStates}
	\begin{align}
	\ket{b_{+}}&=\sin\vartheta\sin\phi\ket{1}+\cos\vartheta\sin\phi\ket{3}+\cos\phi\ket{2},\\
	\ket{b_{-}}&=\sin\vartheta\cos\phi\ket{1}+\cos\vartheta\cos\phi\ket{3}-\sin\phi\ket{2},\\
	\label{eqDarkState}
	\ket{d}&=\cos\vartheta\ket{1}-\sin\vartheta\ket{3},
	\end{align}
\end{subequations}
where the two mixing angles are given by
\begin{subequations}
	\begin{align}
	\label{eqMixingAngleTheta}
	\vartheta(t)&=\arctan\frac{\rabi{P}(t)}{\rabi{S}(t)},\\
	\label{eqMixingAnglePhi}
	\phi(t)&=\frac{1}{2}\arctan\frac{\rabi{rms}(t)}{\Delta}
	\end{align}
\end{subequations}
with the root mean square Rabi frequency ${\rabi{rms}(t)=\sqrt{|\rabi{P}(t)|^2+|\rabi{S}(t)|^2}}$.
STIRAP requires pump and Stokes pulses in the so-called counter-intuitive order, as shown in Fig. \ref{figTheoryResVsDetRegime}(b), when the Stokes pulse precedes the pump pulse by a time delay $\tau$.
We term this a SP pulse pair (We note, that if the system is initially in state $\ket{3}$, adiabatic passage to state $\ket{1}$ requires a reversed pulse order, i.e., a PS pulse pair).
STIRAP transfers the population completely from state $\ket{1}$ to state $\ket{3}$ via the dark state $\ket{d}$, without (ideally) populating state $\ket{2}$.
The dynamics are mirrored by the evolution of the mixing angle $\vartheta$. As the SP pair changes the latter from $\vartheta=0$ to $\vartheta=\pi/2$, the dark state evolves from $\ket{d}=\ket{1}$ to $\ket{d}=-\ket{3}$.
During the process, we must maintain adiabaticity, i.e., the system must remain in the dark state at all times. This requires $\rabi{rms}(t)\gg |\dot{\vartheta}|$.
For smooth pulses (e.g., with Gaussian temporal shape) this adiabatic condition transforms to the simpler form $\mathcal{A}=\int\rabi{rms}(t)\text{d}t \gg 1$, i.e., the pulse area $\mathcal{A}$ has to be sufficiently large \cite{Vitanov2017}.
The larger $\mathcal{A}$, the closer the transfer efficiency approaches unity.
Hence, under realistic conditions of limited pulse area (i.e., finite interaction time and limited pulse intensity) the fidelity of the transfer efficiency of STIRAP is always limited.
This becomes a particular obstacle, if repeated (cyclic) STIRAP processes are required \cite{Beil2011}.

\subsection{Composite pulses\label{subsecUCP}}

Composite pulses replace a single excitation pulse by a sequence of pulses with appropriately chosen relative phases. The latter serve as control parameters to choose an optimized excitation path in Hilbert space, which increases fidelity and robustness with respect to certain errors \cite{Levitt1986}. Composite pulses were so far mainly applied in two-state systems. 
To explain the basic concept in simple terms, we consider now a two-level system rather than the general three-level system required for STIRAP. Nevertheless, the composite approach can be transferred straightforwardly to a three-level scheme \cite{Torosov2013}. As an example, we briefly summarize now the derivation of universal composite pulses for population inversion (sequences termed U5a and U5b), which we also applied in our experiments discussed below. The theoretical treatment follows previous work, which gives a detailed derivation \cite{Genov2014}.

Our objective is to achieve complete population inversion in a two-state quantum system even when the properties of the driving pulses are \emph{unknown}. We assume that the composite pulse duration is shorter than the decoherence time of the system, so its evolution due to a single pulse can be characterized by the propagator $\mathbf{U}$, which connects the probability amplitudes at the initial and final times $t_{\text{i}}$ and $t_{\text{f}}$: $\text{c}(t_{\text{f}})=\mathbf{U} \text{c}(t_{\i})$.
It is conveniently parameterized by
\begin{equation} \label{2stateU}
\mathbf{U} = \begin{pmatrix} \epsilon \e^{\mi\alpha}  & \sqrt{1-\epsilon^2} \e^{\mi\beta} \\  -\sqrt{1-\epsilon^2} \e^{-\mi\beta} & \epsilon \e^{-\mi\alpha} \end{pmatrix},
\end{equation}
where the phases $\alpha$ and $\beta$ and an error term $\epsilon\in[0,1]$ are unknown. Then, the transition probability of the single pulse is $P^{(1)}=1-|U_{11}|^2= 1-\epsilon^2$.
A constant phase shift $\varphi$ in the Rabi frequency leads to $\beta\to\beta+\varphi$ in the propagator $\mathbf{U}(\varphi)$. Then, the propagator of a composite sequence of $N$ identical pulses, each with a phase $\varphi_{k}$, reads $\mathbf{U}^{(N)}= \mathbf{U} (\varphi_{n})\cdots\mathbf{U}(\varphi_{2})\mathbf{U}(\varphi_{1})$.
We make no assumptions about the individual pulses in the composite sequence, i.e., how $\epsilon$, $\alpha$ and $\beta$ depend on the interaction parameters.
This justifies the term ``universal'' for these composite pulses because they will compensate imperfections in \emph{any} interaction parameter. We only assume that the constituent pulses are identical and that we can control their phases $\varphi_{k}$.

In order to determine the phases $\varphi_{k}$, we analyze the propagator element $U^{(N)}_{11}$. It proves useful to choose $\varphi_{k}=\varphi_{N-k+1}$ and we take $\varphi_1=0$ without loss of generality. Thus, in case of a five-pulse sequence ($N=5$), we obtain
\begin{align}
U_{11}^{(5)} &=\left\{[1+2\cos{(2\varphi_2-\varphi_3)}] \e^{\mi\alpha}\right.\notag\\
&+ \left. 2\cos{(\varphi_2-\varphi_3)} \e^{-\mi\alpha}\right\}\epsilon + O(\epsilon^3).
\end{align}
The first order error term vanishes for two distinct sets of phases:
$(\varphi_2=5\pi/6$, $\varphi_3=\pi/3)$ and $(\varphi_2=11\pi/6$, $\varphi_3=\pi/3)$, corresponding to the U5a and U5b composite sequences \cite{Genov2014}.
As the error term $\epsilon$ is typically small, the composite pulse transition probability $P^{(5)} = 1 - O(\epsilon^6)$ is much closer to unity than the transition probability of a single pulse $P^{(1)} = 1 -\epsilon^2$.

In the following, we shift our attention back to the three-state system, where we combine composite pulses and STIRAP to achieve efficient and robust population transfer. 
As theoretically proposed in \cite{Torosov2013}, we describe separately the cases for resonant and highly detuned CSTIRAP.

\subsection{Resonant CSTIRAP\label{subSecResTheory}}

We first consider the case of both lasers tuned to single-photon resonance ($\detuning{P/S}=0$). 
Then, STIRAP is quite sensitive with regard to a proper preparation of the initial state. If the latter is not perfectly aligned with the dark state $\ket{d}$, the obtained transfer efficiency varies strongly with the initial population distribution between states $\ket{1}$ and $\ket{3}$.
Specifically, if a non-negligible fraction of the population is initially placed in state $\ket{3}$, a resonant SP pulse pair will transfer this population adiabatically via the bright states $\ket{b_{\pm}}$. Then, the interference between the two excitation paths (via $\ket{b_{+}}$ and $\ket{b_{-}}$) leads to generalized Rabi oscillations and the transfer efficiency becomes highly sensitive to the pulse area $\mathcal{A}$ \cite{Vitanov2017}. 
This makes resonant STIRAP unsuitable for robust inversion of unknown states.
Additionally, even if all population is initially in the dark state $\ket{d}$, the efficiency of a single STIRAP suffers from residual diabatic losses due to limited adiabaticity. Then, the total efficiency for repeated STIRAP drops quickly as every subsequent inversion is increasingly performed by (the highly sensitive) population transfer via the bright states \cite{Beil2011}.

Perfect adiabaticity in STIRAP can only be reached asymptotically in the limit of infinitely large pulse areas. In order to improve the STIRAP efficiency and robustness also for limited adiabaticity, Torosov \textit{et al.} recently proposed a composite version of STIRAP, i.e., CSTIRAP \cite{Torosov2013}. A sequence of $N$ STIRAP pulse pairs is applied to transfer the population back and forth between initial and target state.
The phases $\varphi_\text{P/S}$ of the individual pump and Stokes pulses serve as control parameters to reduce infidelities in the single STIRAP transfer processes. In order to ensure that the initial state for each STIRAP cycle remains closely aligned with the dark state, the pulse ordering alternates for each pulse pair. Hence, each cycle drives STIRAP with the pulse ordering matched to the transfer direction. Figure \ref{figTheoryResVsDetRegime}(c) shows an example for the pulse sequence and phases for resonant CSTIRAP with three pulse pairs (termed resonant sequence R3). Phases of resonant CSTIRAP with three (R3) and five (R5) STIRAP pulse pairs are compared in Table \ref{tabTheoryPhases}. An analytical expression for the individual pump and Stokes phases for any odd number of pulse pairs is derived in \cite{Torosov2013}. The authors of the theory proposal numerically investigated the performance of resonant CSTIRAP and found that CSTIRAP is expected to outperform conventional STIRAP in terms of peak transfer efficiency and robustness.

\subsection{Detuned CSTIRAP\label{subsecSystem}}
We now consider the case of detuned STIRAP, i.e., when the single-photon detuning ${\Delta\gg\rabi{P/S}}$ is large, while two-photon resonance $\delta=0$ is maintained. For the theoretical description, we can adiabatically eliminate the excited state $\ket{2}$ from the system (see Figure \ref{figTheoryResVsDetRegime}(d)). This transforms the three-level system into an effective two-level scheme.
The pump and Stokes pulses couple the two remaining ground states $\ket{1}$ and $\ket{3}$ with an effective two-photon Rabi frequency ${\rabi{E}(t)=-\rabi{P}(t)\rabi{S}(t)/(2\Delta)}$ and an effective detuning ${\detuning{E}(t)=(|\rabi{P}(t)|^2-|\rabi{S}(t)|^2)/(2\Delta)}$. Figure \ref{figTheoryResVsDetRegime}(e) shows the temporal behavior of the Rabi frequency and detuning after adiabatic elimination for a SP pulse pair on two-photon resonance. The symmetric temporal change of the detuning $\detuning{E}(t)$ over the resonance resembles rapid adiabatic passage (RAP) in the effective two-state system \cite{Vitanov2001}.

While adiabatic population transfer in STIRAP (driven by a SP pulse pair) goes via the dark state $\ket{d}$, in detuned STIRAP also the reversed pulse ordering (i.e., a PS pulse pair) enables smooth, adiabatic transfer. In the latter case, the transfer goes via one of the bright states $\ket{b_{\pm}}$, depending on the sign of the detuning. Therefore, this version of detuned STIRAP, driven by a PS pulse pair, is termed bright STIRAP (b-STIRAP) \cite{Klein2008}. In contrast to standard STIRAP via the dark state, the (decaying) intermediate state $\ket{2}$ can be transiently populated during b-STIRAP, and losses due to radiative decay may occur. However, the amount of transient population in state $\ket{2}$ is negligible for large detuning. Then, a SP or a PS pulse pair can induce a population transfer with (approximately) equal efficiency. Thus, detuned STIRAP is insensitive to the initial state, which makes the technique well suited for inversion of unknown states.

\begin{figure*}
	\includegraphics[width=\textwidth]{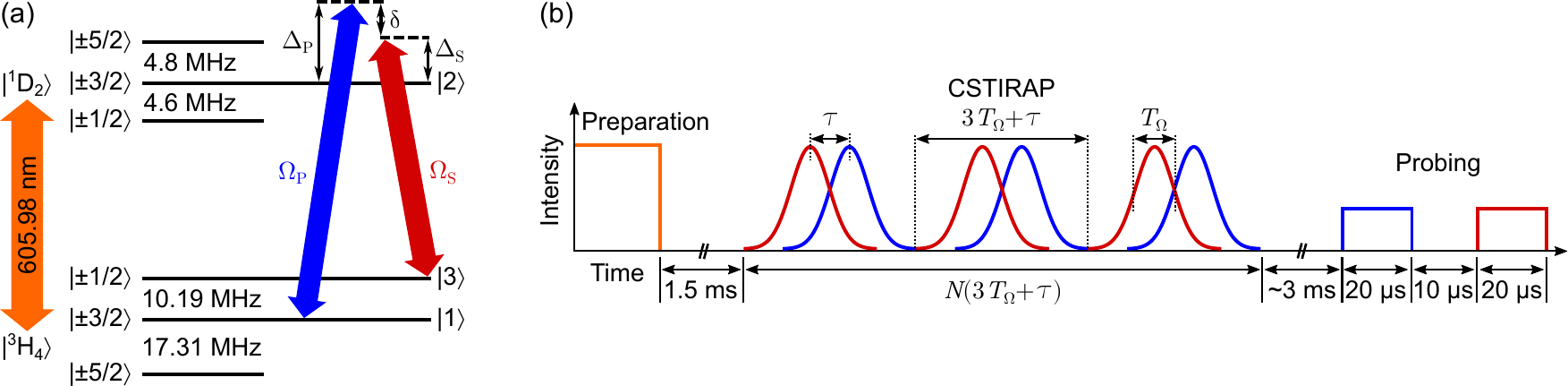}
	\caption{(a) Level scheme of the relevant transitions in \prysoshort. (b) Temporal pulse sequence for preparation, STIRAP (or CSTIRAP) and probing. As an example, we depict a CSTIRAP sequence of three pulse pairs with a non-alternating order. The Rabi frequencies of all applied pulses have a temporal Gaussian shape with a duration (FWHM) of $T_\Omega=\SI{17}{\micro\second}$ in the resonant and $T_\Omega=\SI{14}{\micro\second}$ in the detuned case. We truncate the pulses at $3T_\Omega$. There is no delay between consecutive pulse pairs in a sequence, in order to minimize the duration of the total sequence. The pulse delay $\tau$ between pump and Stokes pulses in each pair is systematically varied in the experiments.}
	\label{figLevelschemeAndSeq}
\end{figure*}

Similarly to the resonant case, the performance of detuned STIRAP depends on the fulfillment of the adiabaticity condition. Perfect adiabaticity in the effective two-state system (see Fig. \ref{figTheoryResVsDetRegime}(e)) can only be reached asymptotically in the limit of infinitely large pulse areas $\mathcal{A}_{\text{E}}= \int|\Omega_{\text{E}}(t)|\text{d}t$ \cite{Vitanov2001}. In order to improve the performance, Torosov \textit{et al.} also theoretically proposed a detuned CSTIRAP version \cite{Torosov2013} and derived an analytical expression for the phases for any odd number of pulse pairs. We note, that the solutions for the phases are very different from the resonant version and the pulse pairs have a non-alternating ordering in detuned CSTIRAP (see Fig. \ref{figTheoryResVsDetRegime}(f)). The latter also ensures that the same pulse characteristics are repeated (up to a phase shift) for every transfer process in the effective two-state system after adiabatic elimination.
Then, only the relative phase $\varphi_\text{P}-\varphi_\text{S}$ between the pump and Stokes fields is important as it determines the phase of $\rabi{E}(t)$ in the effective two-state system. Therefore we can choose $\varphi_\text{S}=0$ without loss of generality and only use $\varphi_\text{P}$ as a control parameter.
Figure \ref{figTheoryResVsDetRegime}(f) shows an example for a pulse sequence and phases for detuned CSTIRAP with three non-alternating pulse pairs (termed detuned sequence D3), as proposed in \cite{Torosov2013}. The phases of detuned CSTIRAP sequences with three (D3) and five (D5) pulse pairs are given in Table \ref{tabTheoryPhases}. Torosov \textit{et al.} numerically investigated the performance of detuned CSTIRAP and confirmed its improved performance compared to detuned conventional STIRAP.
\begin{table}[t]
	\centering
	\begin{tabular*}{\columnwidth}{c @{\extracolsep{\fill}} l l}
		\hline\hline	
		Sequence & $\varphi_\text{P}$ & $\varphi_\text{S}$ \\
		\hline
		R3 & $(0,3,1)\pi/3$&$(1,3,0)\pi/3$\\
		R5 & $(0,5,3,8,4)\pi/5$&$(4,8,3,5,0)\pi/3$\\
		\hline
		D3 & $(0,1,0)2\pi/3$& $(0,0,0)$ \\
		D5 & $(0,2,1,2,0)2\pi/5$& $(0,0,0,0,0)$ \\	
		\hline
		U3 & $(0,1,0)\pi/2$ & $(0,0,0)$ \\
		U5a & $(0,5,2,5,0)\pi/6$ & $(0,0,0,0,0)$ \\ 	
		U5b & $(0,11,2,11,0)\pi/6$& $(0,0,0,0,0)$ \\
		\hline\hline
	\end{tabular*}
	\caption{Calculated phases for different CSTIRAP sequences with three and five pulse pairs. $\varphi_\text{P}$ and $\varphi_\text{S}$ are the phases of the single pump and Stokes pulses. Sequences labeled ``R'' correspond to resonant CSTIRAP. Sequences labeled ``D'' correspond to detuned CSTIRAP, as originally proposed by \cite{Torosov2013}. Sequences with a label ``U'' correspond to our universal version of detuned CSTIRAP. The number attached to the labels denotes the number $N$ of pulse pairs in a sequence. U5a and U5b are two versions of universal detuned CSTIRAP (see text).}
	\label{tabTheoryPhases}
\end{table}

We note, that the solutions for detuned CSTIRAP in \cite{Torosov2013} are designed to improve the performance for limited (weak) adiabaticity due to insufficiently large pulse areas. However, the sequences do not compensate other errors or variations, e.g., in the two-photon detuning $\delta$. It is well known that the efficiency of STIRAP is highly sensitive to the latter (hence, to laser frequency changes or two-photon inhomogeneous broadening) \cite{Vitanov2017}.
Recently, we theoretically derived and experimentally demonstrated \emph{universal} composite pulse sequences for excitations in two-level systems, which compensate against any kind of pulse error and for any arbitrary temporal pulse shape \cite{Genov2014} (see also Sec. \ref{subsecUCP}).
As detuned STIRAP effectively uses a two-level system (see Fig. \ref{figTheoryResVsDetRegime}(d)), we propose now to further improve CSTIRAP by making use of the phases of the universal composite pulses rather than the original phases from \cite{Torosov2013}. We will term this new variant universal detuned CSTIRAP. It enables us to compensate any (repeated) pulse error, e.g., also variation in the driving laser frequencies. Examples for the phases of the pump and Stokes fields for universal detuned CSTIRAP with three (U3) and five (U5a and U5b) pulse pairs are given in Table \ref{tabTheoryPhases}. Phases for higher order universal composite pulses were also derived in our previous work \cite{Genov2014}.

In the following we will present the experimental implementation of resonant CSTIRAP, detuned CSTIRAP, and universal detuned CSTIRAP. We investigated the efficiency and robustness of the composite sequences, and compared the results with single and repeated STIRAP.

\section{Experimental Setup\label{secExpSetup}}

We apply (C)STIRAP for population transfer between two hyperfine ground states of Praseodymium ions doped into an Yttrium orthosilicate crystal (\pryso, hereafter termed \prysoshort), with dimensions of \SI[product-units=single]{5x5x3}{\milli\meter} and a dopant concentration of \SI{0.05}{\atpercent}. The \prysoshort\ crystal is mounted inside a continuous flow cryostat (Janis ST-100), where it is cooled to temperatures below \SI{4}{\kelvin} to suppress phononic excitations. Figure \ref{figLevelschemeAndSeq}(a) shows the relevant level scheme of the $\text{Pr}^{3+}$ ions. In the environment of the host crystal, the two electronic states $\ket{^3\text{H}_4}$ and $\ket{^1\text{D}_2}$ each split up into three hyperfine levels. The population lifetime of the excited states is $T_1^\text{opt}\approx \SI{164}{\micro\second}$. The coherence lifetime of the ground state transitions is $T_2^\text{HF}\approx\SI{500}{\micro\second}$. We note that the coherence time can be increased up to the order of $T_2^{\text{HF}}\approx 1$ s by applying a static magnetic field to prepare appropriate, less sensitive level splitting of the hyperfine ground states \cite{Heinze2014}.
However, this was not necessary in our experiment.
The ultimate limit is set by the spin relaxation time of the hyperfine ground states in \prysoshort, which is of the order of $T_1^\text{HF}\approx 100$ s. Inhomogeneities in the host crystals lattice lead to different transition frequencies for different dopant ions. This gives rise to an inhomogeneous broadening of the optical transition ($\Gamma^\text{opt}_\text{inh}\approx\SI{7}{\giga\hertz}$) and the hyperfine transitions ($\Gamma^\text{HF}_\text{inh}\approx\SI{30}{\kilo\hertz}$). We apply an optical pumping sequence prior to all (C)STIRAP \footnote{We use the notation (C)STIRAP whenever a statement refers to both, STIRAP and CSTIRAP.} measurements to isolate a $\Lambda$-system as shown in Fig. \ref{figLevelschemeAndSeq}(a) from the inhomogeneous manifold. For details on the preparation sequence see \cite{Beil2008}.

The required optical radiation at a wavelength of ${\lambda=\SI{605.98}{\nano\meter}}$ is provided by a solid state laser system \cite{Mieth2014} which is stabilized to a frequency jitter below \SI{100}{\kilo\hertz} on a timescale of \SI{100}{\milli\second}. A small fraction \SI{<1}{\percent} of the light is used as a probe beam, while the remaining radiation is equally split into two beam lines to serve as pump and Stokes beams. All three beams propagate through acousto-optical modulators (AOMs) (Brimrose BRI-TEF-80-50-.606) in double pass configuration, to provide laser pulses with appropriate center frequency and temporal intensity pulse profile. The AOMs in the pump and Stokes beam also control the relative phases of the beams. Using an arbitrary waveform generator (Tektronix AWG 5014B) to drive the AOMs, the setup achieves a phase accuracy of $\SI{<0.5}{\degree}$ with a phase jitter of roughly \SI{0.7}{\degree} on a timescale of \SI{100}{\micro\second}, which is the duration of a typical CSTIRAP sequence. The probe beam is focused inside the crystal to a diameter (full width at half maximum, FWHM) of $\SI{190}{\micro\meter}$. It is overlapped with the pump and Stokes beams, which are collimated to slightly elliptical shapes with dimensions (FWHM) \SI{600x380}{\micro\meter} (pump) and \SI{480x400}{\micro\meter} (Stokes), resulting in peak Rabi frequencies of $\rabi{P/S}^\text{max}\approx 2\pi\times\SI{700}{\kilo\hertz}$ which we estimated by monitoring the transmission of a weak probe field while simultaneously driving Rabi oscillations on the same transition. The diameters of the pump and Stokes beams are chosen much larger than the probe focus to assure rather uniform pump and Stokes Rabi frequencies in the probed volume.

Figure \ref{figLevelschemeAndSeq}(b) shows a typical measurement sequence. We prepare the system in state $\ket{1}$ while state $\ket{3}$ is fully emptied, i.e., the initial populations are $P_1^\text{ini}=1$ and $P_3^\text{ini}=0$. We apply a (C)STIRAP sequence, which transfers population from state $\ket{1}$ to state $\ket{3}$. The sequence duration is well below the coherence time $T_2^\text{HF}$ to ensure a proper phase relationship between the pulses and the single pulse FWHM duration $T_\Omega$ is shorter than $T_1^\text{opt}$ to minimize decay losses for resonant (C)STIRAP \cite{Torosov2013}. After (C)STIRAP we determine the final populations $P_1^\text{final}$ and $P_3^\text{final}$ by absorption measurements with two weak probe pulses at the pump and the Stokes transition. The measured absorption coefficients $\alpha_{i2}$ are related to the populations via $\alpha_{i2} \propto f_{i2} P_{i}$, with the oscillator strengths $f_{ij}$ known from literature \cite{Nilsson2004} (and also confirmed in our own spectroscopic measurements). This permits determination of the transfer efficiency as $\eta = P_3 = (1+x)^{-1}$ with $x=(\alpha_{12}/\alpha_{32})(f_{32}/f_{12})$.
In principle it should be sufficient to probe only the population transferred to the target state $\ket{3}$, as losses outside our three-level system are negligible. However, at large transfer efficiency the absorption on the Stokes transition is very strong. In this case, absorption measurements suffer from a low signal-to-noise ratio. Probing both final populations $P_1^\text{final}$ and $P_3^\text{final}$ overcomes this problem, as there is always strong absorption on one and weak absorption on the other transition. This yields a significantly improved accuracy of the measured absorption coefficients and, hence, the transfer efficiency. With the double-probe approach, we estimate the uncertainty in the transfer efficiency well below \SI{3}{\percent} (for the regime of large transfer efficiency). This is much improved, e.g, compared to the first demonstration of STIRAP in \prysoshort\ with a single probe pulse, yielding uncertainties up to \SI{20}{\percent} for large transfer efficiency \cite{Klein2008}. We note, that the probe pulses are delayed by more than \SI{3}{\milli\second} with respect to the (C)STIRAP sequence, i.e., much longer than the lifetime $T_1^\text{opt}$ of the excited state. Thus, the small fraction of population left due to residual diabatic couplings during (C)STIRAP in state $\ket{2}$ decays back to the ground states. This slightly affects the determination of the obtained transfer efficiency. For efficient transfer by (C)STIRAP the residual population in state $\ket{2}$ will be very small and the error in the measured transfer efficiency will be negligible.
In the worst case of, e.g., intuitive pulse ordering in the resonant regime, we estimate a maximal error in the range of 16 \% compared to the measured transfer efficiency.
However, these high errors are present only for very inefficient population transfer and high residual population in state $\ket{2}$.
Our numerical simulations, which do not take incoherent decay after the pulses into account, nevertheless fit very well to the experimental data. This confirms, that the measured transfer efficiencies are accurate.

\section{Experimental Results\label{secExpRes}}

\subsection{Resonant CSTIRAP\label{subSecResResults}}
We start our investigations with the implementation of resonant CSTIRAP, as theoretically proposed by Torosov \textit{et al.} \cite{Torosov2013}. Thus, the pump and Stokes frequencies are matched to the corresponding transition frequencies. We apply the measurement sequence shown in Figure \ref{figLevelschemeAndSeq}(b) and systematically vary the time delay between the pump and Stokes pulses in each pulse pair of the (C)STIRAP\@ sequence.

\begin{figure}
	\centering
	\includegraphics[width=\columnwidth]{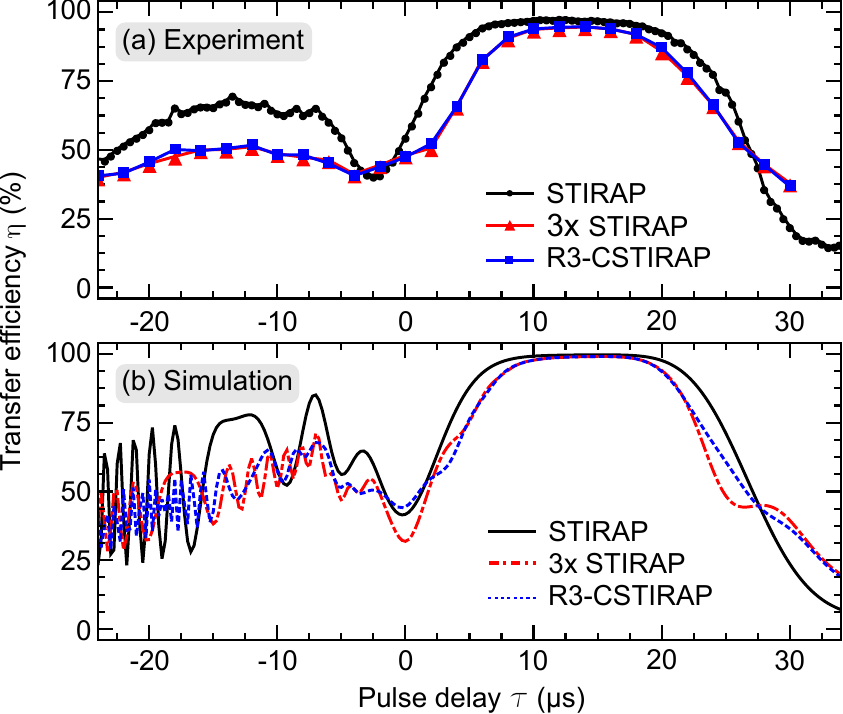}
	\caption{Experimental data (a) and numerical simulations (b) for transfer efficiencies using resonant (C)STIRAP processes vs. variation of the pulse delay between the pump and Stokes pulses. The peak Rabi frequencies are $\rabi{P}\approx 2\pi\times\SI{635}{\kilo\hertz}$ and $\rabi{S}\approx 2\pi\times\SI{510}{\kilo\hertz}$.}
	\label{figResExpAndSim}
\end{figure}

Figure \ref{figResExpAndSim}(a) shows the obtained transfer efficiencies for single STIRAP\@ (single pulse pair), three repeated STIRAPs (with alternating pulse ordering), and the R3-CSTIRAP sequence (with alternating pulse ordering) vs. the pulse delay $\tau$ between the pump and Stokes pulses. In the graph positive delays $\tau>0$ correspond to the counter-intuitive pulse ordering (Stokes preceding pump), while negative delays $\tau<0$ correspond to the intuitive pulse ordering (pump preceding Stokes). The pulse areas are $\mathcal{A}\approx 20\pi \gg 1$ for a single pulse pair, i.e., we fulfill the adiabatic condition well. Hence, for counter-intuitive pulse ordering $\tau>0$, STIRAP yields a broad plateau of robust and efficient transfer with a peak efficiency of \SI{97}{\percent}. Intuitive pulse ordering $\tau<0$ yields much smaller peak efficiencies, as expected from theory. We note, that at intuitive pulse ordering $\tau<0$ we might expect to observe pronounced oscillations of the transfer efficiency due to 
interference between the excitation paths via the two bright states \cite{Vitanov2017}.
However, the fast oscillations are washed out due to averaging over spatially varying Rabi frequencies across the laser profiles. This averaging is irrelevant for STIRAP (i.e., the excitation dynamics for $\tau>0$), as long as the adiabatic condition is fulfilled. The repeated resonant STIRAP yields systematically reduced efficiencies in comparison to single STIRAP, as the errors of the imperfect three STIRAP processes accumulate, and each transfer process leaves the system in a less pure initial state for the next STIRAP cycle. When we apply R3-CSTIRAP, the composite version has a lower efficiency than STIRAP, contrary to what we would expect from simple theory. Moreover, the R3-CSTIRAP sequence also does not show a measurable improvement compared to repeated conventional STIRAP.

\begin{figure*}
	\centering
	\includegraphics[width=\textwidth]{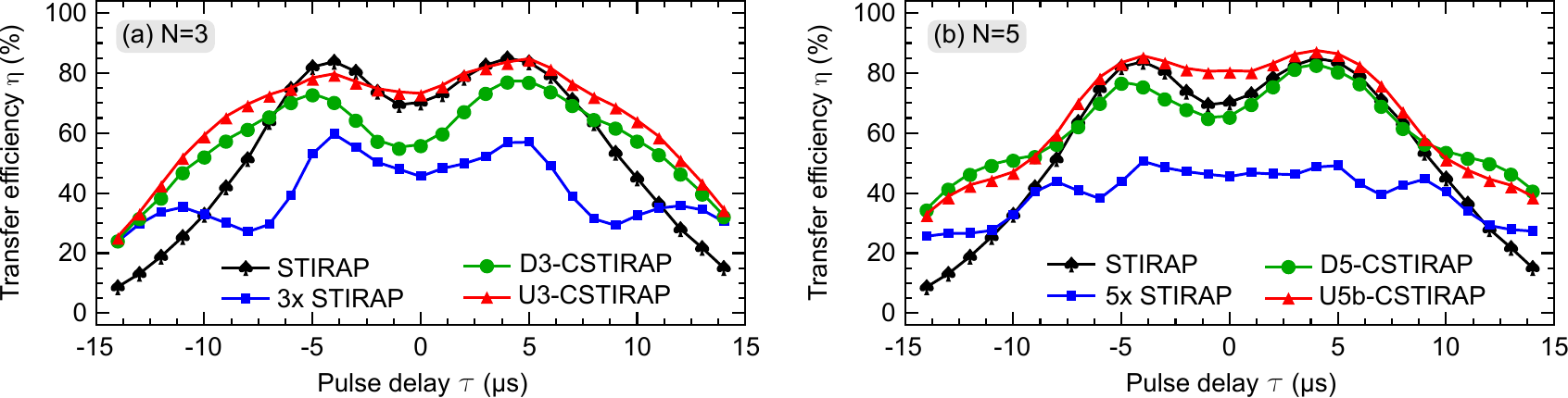}
	\caption{Experimentally determined transfer efficiencies for detuned (C)STIRAP processes vs. variation of the pulse delay between the pump and Stokes pulses. We apply (a) $N=3$ or (b) $N=5$ pulses in the repeated (C)STIRAP sequences. The single pulse duration is $T_\Omega=\SI{14}{\micro\second}$ and the single-photon peak Rabi frequencies are $\rabi{P}=2\pi\times\SI{640}{\kilo\hertz}$ and $\rabi{S}=2\pi\times\SI{550}{\kilo\hertz}$.}
	\label{figExpDetOptimized1DAll}
\end{figure*}

This lack of improvement by R3-CSTIRAP is mainly due to optical inhomogeneous broadenings in \prysoshort. Specifically, resonant CSTIRAP is designed to compensate weak adiabaticity but remains sensitive to single-photon inhomogeneous broadening. This is due to its error compensating mechanism, which relies on a symmetry of the Hamiltonian that is present only when the pump and Stokes frequencies are tuned to single photon resonance $\Delta=0$ \cite{Torosov2013}. However, the \prysoshort\ medium exhibits residual inhomogeneous broadenings of the optical transitions in the range of \SIrange{200}{300}{\kilo\hertz} after optical preparation. Moreover, also instantaneous spectral diffusion yields additional broadenings whenever the intermediate state is populated \cite{Huang1989}. The pump and Stokes Rabi frequencies in the CSTIRAP sequence have a temporal Gaussian shape with a duration (FWHM) of $T_\Omega = \SI{17}{\micro\second}$. Hence, the pulse bandwidth is not sufficiently large to cover the optical inhomogeneous broadening, which would require $1/T_\Omega\gg\Delta$ \cite{Torosov2013}.
Hence, CSTIRAP also enables large transfer efficiency, but does not exceed the performance of STIRAP.

We confirmed these arguments by numerical simulations where we assumed Gaussian-shaped inhomogeneous broadenings of the optical (bandwidth $\Gamma^\text{opt}_\text{inh}$ = \SI{200}{\kilo\hertz} (FWHM)) and the hyperfine (bandwidth $\Gamma^\text{HF}_\text{inh}$ = \SI{30}{\kilo\hertz} (FWHM)) transition as well as excited state decay time $T_1^\text{opt}$ and the hyperfine transitions decoherence time $T_2^\text{HF}$.
We simulated the dynamics of the three-level system with the density matrix formalism by solving the Liouville-von Neumann equation, where optical decay and spin decoherence rates are included as imaginary elements of the Hamiltonian \cite{Ivanov2004}. In order to take into account optical and spin inhomogeneous broadening, we perform each simulation for 961 atoms with different single-photon (optical) detunings in the range between \SI{\pm 300}{\kilo\hertz} with a step of \SI{20}{\kilo\hertz} and two-photon (spin) detunings in the range between \SI{\pm 60}{\kilo\hertz} with a step of \SI{4}{\kilo\hertz}. The density matrix of the atomic ensemble is then calculated as a weighted average of the density matrices of the individual atoms, taking into account the probability distribution of the optical and spin inhomogeneous broadenings. We estimated the effect of optical, spin inhomogeneous broadening, optical decay, and spin decoherence by turning them on and off in the simulation and calculating the effect of each factor on the transfer efficiency of (C)STIRAP for the ensemble.
In the simulation we do not take into account decay after (C)STIRAP in order to consider only the transfer efficiency due to (C)STIRAP. The simulated transfer efficiencies are shown in Fig. \ref{figResExpAndSim}(b). As already mentioned, the fast oscillations for $\tau<0$ mirror diabatic excitation dynamics. The oscillations are washed out in the experiment due to spatial averaging, which was also confirmed numerically.
Nevertheless, the simulations involving inhomogeneous broadenings fit very well with the experimental data, even with particular details such as the maximal transfer efficiencies, the extension of the plateaus, or the averaged efficiency for $\tau<0$, or the overall lineshape. The simulations clearly confirm, that resonant CSTIRAP in the optically inhomogeneously broadened medium reaches towards, but cannot exceed the efficiency or robustness of STIRAP.

In order to overcome the optical inhomogeneous broadenings, we tried to increase the bandwidth of the resonant (C)STIRAP pulses by reducing the pulse duration. The numerical simulations indicate, that pulse durations below \SI{1}{\micro\second} would be necessary to achieve a measurable improvement of CSTIRAP over repeated STIRAP. This would be technically possible, though at the limits of our optical setup. However, at this much shorter pulse duration we also had to increase the pulse intensity substantially in order to maintain a sufficiently large pulse area. In this case, off-resonant couplings to other transitions in our medium, outside the three-level system start to play a role. This leads to perturbations in the (C)STIRAP dynamics and severe pulse distortions during propagation through the medium. Detuned (C)STIRAP offers an alternative solution, as it is insensitive to inhomogeneous broadenings, provided the applied detuning is sufficiently large.

\subsection{Detuned CSTIRAP}
For detuned CSTIRAP, the frequencies of the pump and Stokes fields are shifted by a single-photon detuning $\Delta\neq 0$ from the corresponding resonances. In principle, the sign of the detuning does not matter. In our specific experiment the pump and Stokes frequencies are both blue shifted, which leads to only negligible off-resonant excitations to other states outside our specific three-level system in \prysoshort. The detuning should be sufficiently large in order to convert the three-level scheme into an effective two-level Raman-type system. This requires ${\Delta\gg\rabi{P/S}}$. Experimentally, a useful indicator for this case is the observation of equal transfer efficiencies for excitation with a SP pulse pair or a PS pulse pair. At sufficiently large detuning, the pulse ordering plays no role, as the population is completely transferred either by STIRAP or b-STIRAP. On the other hand, the detuning should not be too large, as the two-photon Rabi frequency $\rabi{E}\propto1/\Delta$ decreases with larger detuning. The Rabi frequency must remain sufficiently large to maintain adiabaticity and, hence, permit efficient and robust transfer.
We note, that also longer pulses improve adiabaticity. However, the pulse duration must remain sufficiently short to cover the inhomogeneous broadening $\Gamma^\text{HF}_\text{inh}\approx \SI{30}{\kilo\hertz}$ of the two-photon hyperfine transition $\ket{1}\leftrightarrow\ket{3}$ between the ground states in \prysoshort.

We performed systematic measurements of the transfer efficiency for conventional STIRAP and b-STIRAP to determine the optimal single-photon detuning to be $\detuning{S}=\SI{1.75}{\mega\hertz}$. Subsequently, we permitted also for a small ($\sim\SI{10}{\kilo\hertz}$) variation in the two-photon detuning $\delta$ by changing only the pump frequency $\detuning{P}=\delta+\SI{1.75}{\mega\hertz}$. This served as a simple control parameter to match the effective pump and Stokes Rabi frequencies, enabling equal transfer efficiencies for STIRAP and b-STIRAP.  Though equal Rabi frequencies are no strong requirement for (C)STIRAP, it enables comparison with the theory predictions for detuned CSTIRAP \cite{Torosov2013}, which assumed equal transfer efficiencies for SP and PS pulse pairs. Subsequently, we used these optimized detunings for all further experiments.

\begin{figure*}
	\centering
	\includegraphics[width=\textwidth]{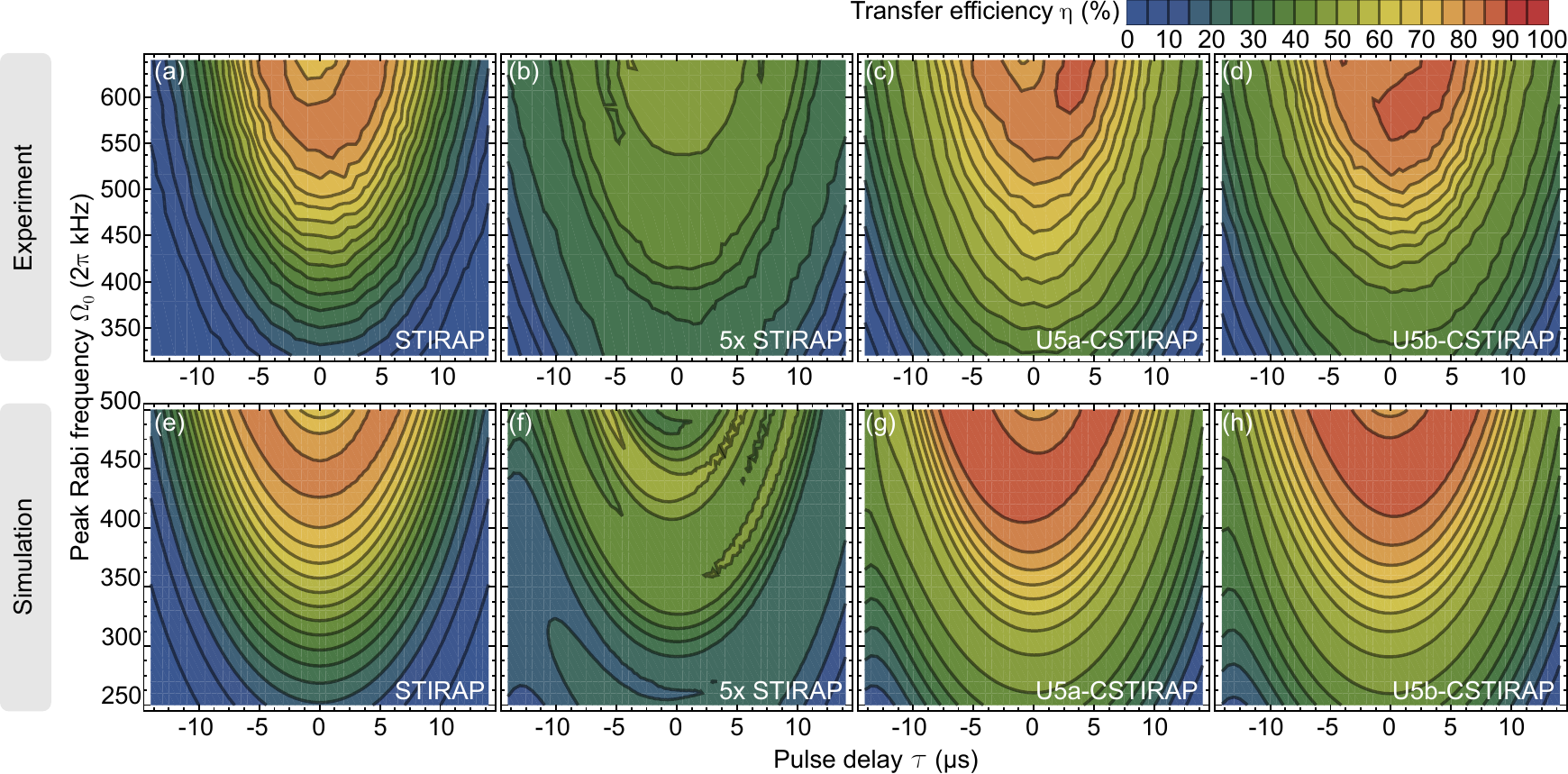}
	\caption{Transfer efficiencies for different detuned (C)STIRAP processes, vs. variation of both pulse delay and peak Rabi frequency. (upper row, (a)-(d)) Experimental data. (lower row, (e)-(h)) Numerical simulations. The single pulse duration is $T_\Omega=\SI{14}{\micro\second}$ and the single-photon peak Rabi frequencies of the pump and Stokes pulses are $\rabi{P}=\Omega_0$ and $\rabi{S}=\num{0.85}\Omega_0$, with $\Omega_0$ given in the plots. Note, that the peak Rabi frequencies used in the simulation systematically differ by roughly \SI{20}{\percent} compared to the experimental values. This is due to spatial field inhomogeneities in the experiment and, hence, averaging effects in the experimental Rabi frequencies. Such spatial averaging is neglected in the simulation.}
	\label{figExpAndSimDetOptimized2DAllExp}
\end{figure*}

We performed systematic measurements to compare the performance of detuned STIRAP and CSTIRAP\@. Figure \ref{figExpDetOptimized1DAll}(a) shows the measured transfer efficiencies vs. the pulse delay $\tau$ for detuned STIRAP (single pulse pair), repeated detuned STIRAP\@ (non-alternating pulse ordering), and two variants of detuned CSTIRAP (non-alternating pulse ordering). Single detuned STIRAP yields equal transfer efficiencies above \SI{80}{\percent} for both intuitive ($\tau\approx\SI{-5}{\micro\second}$) and counter-intuitive pulse ordering ($\tau\approx\SI{5}{\micro\second}$). Compared to the resonant case (see Figure \ref{figResExpAndSim}), the peak transfer efficiencies for single detuned STIRAP are smaller, and also the extension of the regions with high transfer efficiency is smaller. This is due to the lower coupling strength for detuned excitation which yields an effective pulse area of about $\mathcal{A}_\text{E}\approx4\pi$.

The transfer processes suffer substantially from accumulated errors in repeated conventional STIRAP. Specifically, repeated STIRAP with three SP pulse pairs reaches peak efficiencies around \SI{50}{\percent} only, also showing pronounced variations of the transfer efficiency vs. pulse delay. When we apply now the D3-CSTIRAP sequence, as proposed in \cite{Torosov2013}, the peak transfer efficiency substantially increases towards \SI{70}{\percent}. Moreover, also the high transfer region broadens compared to repeated STIRAP. Hence, the choice of appropriate phases in D3-CSTIRAP sequence already strongly improves the robustness of the excitation process compared to repeated STIRAP. However, the D3-CSTIRAP sequence does not yet reach the peak efficiency of \SI{80}{\percent} for single STIRAP - though the robustness with regard to delay variations is already larger. Our new U3-CSTIRAP achieves further improvements: It reaches the efficiency of single STIRAP and further increases the width of the region of efficient transfer.

The advantages (in particular of the universal version) of CSTIRAP become even more obvious, when we apply longer pulse sequences. Figure \ref{figExpDetOptimized1DAll}(b) shows the transfer efficiency vs. the pulse delay $\tau$ for single STIRAP, as well as repeated STIRAP, D5-CSTIRAP \cite{Torosov2013}, and universal U5b-CSTIRAP, the latter sequences with 5 pulse pairs (with non-alternating ordering). In terms of composite pulses, the ability for error correction increases with the number of pulses in the sequence. Figure \ref{figExpDetOptimized1DAll}(b) confirms this expectation: While repeated STIRAP with 5 pulse pairs remains at transfer efficiencies well below \SI{50}{\percent}, D5-CSTIRAP reaches the peak efficiency of single STIRAP at \SI{80}{\percent}, while simultaneously increasing the width of the efficient transfer region. Finally, the universal sequence U5b-CSTIRAP outperforms all other configurations. U5b-CSTIRAP combines peak transfer efficiencies around \SI{85}{\percent}, with similarly large robustness vs. delay variations as D5-CSTIRAP. We note that U5a-CSTIRAP (not shown in Fig. \ref{figExpDetOptimized1DAll}(b)) and U5b-CSTIRAP have very similar performance with a slightly better performance for the U5b version for zero pulse delay. The strong performance of the universal sequences is due to their robustness with regard to variations in any arbitrary experimental parameter, while the D3- and D5-CSTIRAP sequences were designed to compensate for limited adiabaticity only (e.g., induced by variations in the driving laser intensity) \cite{Torosov2013}. We note that we also applied higher order sequences of seven and nine pulse pairs for detuned CSTIRAP but they did not improve performance in comparison to U5b-CSTIRAP. This can be explained by the smaller additional error compensation of the higher order sequences, which cannot make up for the higher number of (low efficiency) STIRAPs and the longer duration of the whole sequence. For example, a sequence with $N=9$ pulse pairs exhibits a total duration of more than \SI{500}{\micro\second}, i.e., longer than the coherence time $T_2^\text{HF}$.

We conducted further systematic investigations on universal detuned CSTIRAP vs. variations in multiple experimental parameters. Figures \ref{figExpAndSimDetOptimized2DAllExp}(a)-(d) show measured transfer efficiencies vs. variation of both the pulse delay $\tau$ and peak Rabi frequencies. We took data for single detuned STIRAP\@, repeated detuned STIRAPs, and two universal variants U5a and U5b of CSTIRAP \cite{Genov2014}. Similar to Fig. \ref{figExpDetOptimized1DAll}, we also observe here, that the efficiency of repeated STIRAP drops considerably all over the parameter range compared to single STIRAP. Both universal sequences U5a- and U5b-CSTIRAP fully recover (and even exceed) the high efficiency of single STIRAP also for a five-fold repeated transfer process. U5a- and U5b-CSTIRAP outperfom conventional STIRAP and repeated STIRAP in terms of peak efficiency and broad bandwidth with regard to variations in pulse delay and Rabi frequencies. The peak transfer efficiency of the CSTIRAP sequences is \SI{87}{\percent}, compared to \SI{50}{\percent} for repeated STIRAP.
A detailed comparison of the two universal CSTIRAP variants phases reveals a slightly increased robustness for U5b-CSTIRAP compared to the U5a version (compare, e.g., the extensions of the high efficiency regions, shaded in red/orange in Figs. \ref{figExpAndSimDetOptimized2DAllExp}(c,d)). This confirms previous work on universal sequences, showing that U5b sequences compensate especially well against detuning errors \cite{Genov2014}. Maintaining the two-photon resonance is crucial for STIRAP, and two-photon detunings are an issue in our medium due to the inhomogeneous broadening of the hyperfine transitions. Finally, we confirmed the experimental findings by numerical simulations (see details given in section \ref{subSecResResults}). Figure \ref{figExpAndSimDetOptimized2DAllExp}(e)-(h) shows the simulation results, which reproduce the experimental behavior well.

\section{Conclusion\label{SecConclusion}}
We experimentally demonstrated and systematically studied several variants of composite STIRAP, i.e., a combination of the concepts of composite pulse sequences with adiabatic passage in a three-level scheme.
In particular, we applied CSTIRAP sequences in a rare-earth ion-doped solid for population transfer between hyperfine ground states. We compared the transfer efficiency and robustness of CSTIRAP with conventional single and repeated STIRAP in the resonant and highly detuned regime. In the resonant case, inhomogeneous broadening of the optical transition perturbed CSTIRAP, as the required single-photon resonance cannot be maintained for all frequency ensembles of dopand ions.
Nevertheless, in the highly detuned regime CSTIRAP significantly boosted the peak transfer efficiency by more than \SI{70}{\percent} compared to repeated STIRAP, and also outperformed single conventional STIRAP.
Moreover, CSTIRAP offered much higher robustness with regard to variations in certain experimental parameters. We also compared CSTIRAP sequences with three or five pulse pairs, proving that longer sequences yield better error compensation and performance. Finally, we developed and demonstrated universal detuned CSTIRAP variants. The universal composite sequences are robust with regard to fluctuations in any arbitrary experimental parameter, outperforming also the  originally proposed CSTIRAP sequences. We confirmed all experimental data by numerical simulations, which reproduce the experimental data well.

Our findings are of relevance for any application, which requires improved fidelity and robustness of STIRAP. They are of particular relevance for applications of repeated STIRAP, where limited transfer efficiencies per cycle quickly add up to perturb adiabatic passage in a highly nonlinear fashion. Even small improvements in the transfer efficiency enable many more STIRAP cycles. As a significant advantage compared to conventional STIRAP, the detuned versions of CSTIRAP effectively invert any arbitrary initial superposition of two ground states. They maintain large efficiency also in case of repeated application, e.g., for rephasing or dynamical decoupling in quantum memories, or in atom optics to achieve a larger momentum transfer and beam deflection.

\begin{acknowledgments}
The authors thank K. Bergmann, D. Schraft and B. W. Shore for valuable discussions. This work is supported by the Alexander von Humboldt Foundation, the Deutsche Forschungsgemeinschaft, and a Career Bridging Grant of Technische Universit{\"a}t Darmstadt. NVV acknowledges support by the Bulgarian Science Fund Grant No. DO02/3 (ERyQSenS).
\end{acknowledgments}

%
\end{document}